\documentclass[10pt, twocolumn]{IEEEtran}

\usepackage{graphicx,eucal,amsfonts}
\usepackage{url}

\begin{document}

\title{Toward Resilient Algorithms and Applications}

\author{
Michael~A.~Heroux,  Sandia National Laboratories\\
}

\date{} %{\today} % Activate to display a given date or no date

\maketitle

\section{Introduction}

Since the early days of supercomputing\footnote{The first Cray-1 (SN1) was delivered to Los Alamos National Laboratory without SECDED memory.  Errors were so frequent that SN2 was scrapped and SN3 was delivered to NCAR with SECDED.}, large-scale computing platforms have been engineered to handle unreliability.  In contrast, algorithms and applications for large-scale systems have generally assumed a fairly simplistic failure model:  The computer is a reliable digital machine, with consistent execution times and infrequent failures.  If failure occurs, recovery can be handled by checkpoint-restart (CPR): occasionally storing a snapshot of application state and restarting from that saved state.

Over the past decade, the high performance computing community has become increasingly concerned that preserving the reliable, digital machine model will become too costly or infeasible~\cite{Miskov-Zivanov:2007:SER:1266366.1266680,Karnik:2004:CSE:1032295.1032595}.  With the push toward exascale computing, this concern has become even greater~\cite{InterAgencyResilienceWorkshop2012}, and we must explore other models and improve algorithms.

In this paper we discuss possibilities for developing new algorithms that are resilient to hard and soft failures.  However, in order to reason about such algorithms, we first need programming models that enable more sophisticated recovery strategies than CPR. 

\section{Four Resilience-Enabling Programming Models}

Algorithm-based fault tolerance has certainly been studied, going back many decades~\cite{huang1984algorithm}, and many algorithms have been developed~\cite{Reis:2005:SSI:1048922.1048991,Langou:2007:RPI:1350656.1350657,Wilfredo:2000:SFT:886626,Yang:2007:FTP:1299042.1299048}, but none of these algorithms have made it into broad practical use because we have no standard programming model support.  In order to develop effective resilient algorithms and applications, we first need programming models that permit us to reason about failure and implement recovery.  Here we present four programming models that we think have strong promise of being useful, ordering them from easiest to hardest to implement in a production system.  Even though these models are not widely deployed today, using them as abstractions for developing new algorithms will provide motivation and guidance for development of both new algorithms and the underlying system software and hardware.

\flushleft\fbox{\parbox{3.4in}{Sandia National Laboratories is a multi-program laboratory managed and operated by Sandia Corporation, a wholly owned subsidiary of Lockheed Martin Corporation, for the U.S. Department of Energy's National Nuclear Security Administration (NNSA) under contract DE-AC04-94AL85000.}}  
\subsection{Skeptical Programming (SkP)}
Almost all algorithm developers assume that their software will execute reliably or fail obviously by halting.  However, if they understand that silent data corruption is possible, they can develop very simple and inexpensive validation tests based on their understanding of the mathematical properties of their algorithms.  For example, many algorithms have global properties of orthogonality, or conservation principles, that are implicitly assumed to be true during the execution of an algorithm.  If these properties were checked occasionally during execution, the cost can be very low and many silent data corruption events could be detected.  Recovery may be as simple as aborting, or may involve rolling back to a previous valid state, or even continuing execution if the error will be damped by subsequent computations.

\subsection{Relaxed Bulk-synchronous Programming (RBSP)}
One of the first impacts of reduced reliability is performance variability.  As low-level system failure rates increase, error detection and correction happen more frequently in the hardware and system software layers.  These events preserve the reliable digital machine model, but introduce variability in execution time.  Many scalable applications are designed under the implicit assumption that equal work implies equal execution time, so that if we balance the work of a parallel application, we should scale well on a parallel computer, even though we synchronize across processors during execution.  Performance variability, when coupled with frequent collective operations, leads to severe limitations in scalability, especially as we go to a million or more processes.

With the introduction of MPI-3~\cite{MPI-3.0}, we now have asynchronous neighborhood and global collectives, enabling a ``relaxed'' bulk-synchronous programming model (RBSP).  Given RBSP capabilities, we are now able to develop new algorithms that can potentially hide latency.

\subsection{Local Failure, Local Recovery (LFLR)}\label{subsect:LFLR}
For parallel applications based on MPI, the current approach to dealing with the loss of a single process is to kill all remaining processes and restart the application.  As we now regularly run on hundreds of thousands to more than a million processors, this approach is not feasible.  Instead a local failure should permit a local recovery\footnote{By local recovery we do not mean that no communication is done in the detection or recovery phases.  Instead we mean that recovery is logically local from an application developer's perspective.  All processes that have valid state are not involved in recovery, except to the extent that they assist in restoring state on the failed process, or take over its workload.}.  

One type of local-failure-local-recovery (LFLR) model permits the user to store specific data {\it persistently} for each MPI process and allows a recovery function to be registered, such that, if a process fails, a new process is started and assigned to the rank of the failed process, and the user's recovery function is called, giving access to the persistent  data of the old process, as well as the neighbors' persistent data.  Using LFLR, we can develop new algorithms for many types of problems.

\subsection{Selective Reliability Programming (SRP)}
The fourth programming model we discuss is Selective Reliability Programming (SRP), which gives the programmer the ability to declare specific data and compute regions to be more reliable that the ``bulk'' reliability of the underlying system (or we can switch the default to be reliable and then selectively be less reliable).  By distinguishing between what needs to be highly reliable or not, we can develop new algorithms that store most data and do most computations with low reliability while retaining the robustness of a fully reliable approach.

Although the costs of high reliability will impact the practicality of some approaches, the details of how reliability is  implemented is not fundamentally important to reasoning about new algorithms.  In some cases, even very expensive approaches such as triple modular redundancy (TMR) can still be much faster than a fully unreliable approach.

\section{Toward Resilient Algorithms}

Resilient algorithms have long been a subject of research.  The above four programming models enable further research and drive co-development of the algorithms and  computing system features that are required to realize resilient applications.  In this section we discuss some of the many possible algorithms that can be developed under the above programming models in order to provide resilience on future systems.

\subsection{Detecting and Responding to Silent Data Corruption}
Skeptical programming can be used to detect silent data corruption such as bit flips, and then determine if the resulting error is ``harmless'' or not.  One example of this kind of algorithm is an implementation of GMRES~\cite{Elliott2013} that detects and, optionally, corrects single bit flips very inexpensively as part of the Arnoldi process.  Many existing ABFT algorithms can be implemented in using a skeptical algorithm programming approach, since the meta data used to recover state can also be used to detect anomalous behavior.

\subsection{Latency-tolerant Algorithms}

One of the most important and effective algorithm research and development strategies we can explore now is latency tolerance.  Many of our scalable algorithms and applications depend on collective operations that, when implemented in a straightforward manner, lead to synchronous global collectives.  On emerging high end platforms, these collectives have become severe performance limiters due to poor scaling of collectives.  The advent of asynchronous collectives gives us new opportunities.  The basic challenge we face as algorithm developers in this situation is finding useful work to do while a collective is completing.  Recent work in pipelining algorithms, for example the p$(l)$-GMRES algorithm~\cite{ghysels2012hiding}, shows that latency hiding by unrolling iterations in a Krylov solver can help restore scalability.  Similar approaches for many algorithms can lead to relatively minor design changes that result in better tolerance of latency and performance variability.

The impact of successfully redesigning algorithms to be latency tolerant is that performance variability on existing systems can be hidden.  But even more importantly, if we can tolerate performance variability due to error detection and correction at the system software and hardware levels, system designers can detect and correct more errors without impacting application scalability, permitting us to extend the viability of the reliable digital machine model.

\subsection{Locally Restarted PDE Computations}

Given the programming features described in Section~\ref{subsect:LFLR}, we have the potential to develop a broad collection of algorithm with local recovery properties.  Examples for differential equations include:
\begin{itemize}
\item Explicit methods:  As shown in~\cite{GokhaleWong2011}, an explicit time-stepping algorithm can be easily implemented to recover locally, given the LFLR features.
\item Implicit methods:  This case is more interesting.  The challenge is to restore a local state that is equivalent up to the truncation error of the PDE.  Several interesting approaches seem promising. 
\item Redundant storage of coarse model:  In order to recover state from a lost process, we could explore storing a coarse model representation on neighboring processes that could be used to boot-strap state recovery upon failure.
\end{itemize}

\subsection{Reliable Outer Iterations}

Many algorithms can be cast in an outer-inner formulation.  For example, a fault-tolerant GMRES variant, as described in~\cite{2012arXiv1206.1390B}, uses reliable computation and storage in the outer iteration and an ``unreliable'' GMRES in the inner iteration.  The result is that most computation and data are in low-reliability mode, leading to presumably cheaper computations.  Because the outer iterations are reliable, the solution returned by the inner solve (if it comes back at all) can be analyzed and used or discarded.  Even if the inner solve answer is not correct, it can still be used with some effect.

\section{Conclusions}

Resilience is a critical requirement for future high-end computing.  In order to effectively develop resilient algorithms and applications, we need robust and usable resilient computing models.  In this paper we have identified four specific models that allow us to reason about and develop new algorithms.  SkP requires nothing more than a change in attitude on the part of the programmer, from trusting that a machine is reliable digital device to being aware that an incorrect computation may occur.  RBSP is already possible with the introduction of MPI 3.0.  LFLR requires more support from the underlying system layers, and requires some kind of support from programming languages and libraries.  The ULFM library~\cite{Bland2013,ULFM} already provides one approach to supporting LFLR.  SRP is the most challenging model, but also firmly addresses one of the biggest challenges we face: silent errors.

Much of the focus of extreme-scale computing is on massive concurrency, which is appropriate.  However, without resilient computing models we face a very real risk of application failure rates that are too high to realize the benefits of future systems.  In addition, any progress we make in resilient algorithms and applications permits us to utilize lower cost systems in general, systems with lower quality interconnect networks, higher bit failure rates and increased node loss.  Resilient algorithms and applications will enable effective extreme scale computing and reduce system costs at other levels.  

%REFERENCES
\bibliographystyle{IEEEtran}
\bibliography{ResAlgs}

\end{document}